\documentclass[a4paper,12pt]{article}
\usepackage[centertags]{ams math}
\usepackage{amssymb}

\usepackage{graphicx}
\usepackage{epsfig}
\usepackage{ulem}
\usepackage[english]{babel}
\usepackage{array}

\usepackage{latexsym}

\usepackage{yfonts}
\usepackage{mathrsfs}
\usepackage[mathcal]{euscript}

\pdfoutput=1
\usepackage{epsfig}
 \usepackage{jheppubm}
\usepackage{mathdots}
\usepackage{MnSymbol}
\usepackage{multirow}

\definecolor{darkraspberry}{rgb}{0.53, 0.15, 0.34}

\definecolor{darkblue}{rgb}{0., 0, 1}

\definecolor{dgreen}{rgb}{0.,0.6,0.}

\newcommand{\nn}{\nonumber}
\newcommand{\be}{\begin{equation}}
\newcommand{\ee}{\end{equation}}
\newcommand{\bea}{\begin{eqnarray}}
\newcommand{\eea}{\end{eqnarray}}

\title{Violation of the Third Law of Thermodynamics by Black Holes,\\ 
Riemann Zeta Function and\\
Bose Gas in Negative Dimensions 
}
\author{Irina Aref'eva and Igor Volovich}

\affiliation{Steklov Mathematical Institute, Russian Academy of Sciences,\\Gubkina str. 8, 119991, Moscow, Russia}
\emailAdd{arefeva@mi-ras.ru}
\emailAdd{volovich@mi-ras.ru}

\abstract{Black holes violate  the third law of thermodynamics in its standard formulation. Schwarzschild black hole entropy is inverse proportional to the square of the temperature $S=1/(16 \pi T^2) $ and tends to infinity 
rather than zero when the temperature goes to zero. We search for quantum statistical models with such exotic  thermodynamic  behaviour. It is shown that the Schwarzschild black hole in $D=4 $ spacetime dimensions  corresponds to a Bose gas  in a space with $d=-\, 4$ {\it negative} spatial  dimensions. The Riemann zeta function is used to define the entropy of the Bose gas in negative dimension. 
The correspondence between  black  holes in higher dimensions  and de Sitter  spacetime with Bose gases is considered.

}

\begin{document}
\maketitle

\section{Introduction}

There is a remarkable analogy between the laws of thermodynamics and the laws of black hole mechanics \cite{BCH,Bekenstein:1973, Hawking:1975vcx,FN,Wald-book}.
However, there is a problem with the third law of thermodynamics. 
\\

 According to the standard formulation of the third law, the entropy $S$ should tend to zero as the  temperature $T$ goes to zero. This law is satisfied by most models of statistical mechanics \cite{LLV}. 
 The entropy of  Schwarzschild black hole is inverse proportional to the square of the temperature $S=1/(16 \pi T^2) $ and tends to infinity 
rather than zero when the temperature goes to zero.
 Therefore, a microscopic model of black hole entropy cannot be an ordinary quantum system obeying the third law of thermodynamics.  \\

The third law of thermodynamics is also violated for the Reissner-Nordstrom, Kerr, Kerr-Newman and some other  black hole solutions. It is violated not only in the standard Planck's formulation discussed above but also in the Nernst formulation. The temperature of extremal black holes is equal to zero, but its entropy depend on the parameters of the black holes, meanwhile, according to the Nernst formulation of the third law, the entropy values at zero temperature should be a universal constant.\\

The third law of the black hole mechanics is formulated as unattainability of the state with the surface gravity equal to zero \cite{BCH,Israel,Wald,Racz,Belgiorno:2002pm,Wreszinski:2007gu}. A possible violation of this formulation is discussed recently \cite{Kehle:2022uvc}.
\\

The  problem  with the microscopic  origin  of the Bekenstein-Hawking entropy \cite{Bekenstein:1973,Hawking:1975vcx} for  black holes  is that black holes do not satisfy the third law of thermodynamics  in its standard  formulation.
Therefore, such exotic thermodynamics behaviour of black hole cannot be obtained by using ordinary quantum statistical mechanics models which obey the third law. 
In this paper we show that the entropy of the   Schwarzschild black hole
\be
S_{BH}=\frac{\beta ^2}{16 \pi G},  \qquad\beta=1/T
\ee
($G$ is the Newton constant) corresponds to the Bose gas in $d=-\,4$ {\it negative} spatial dimensions. This conclusion is obtained by using properties of the Riemann zeta function. The entropy
of the Bose gas in $d$-dimensional space, $d=1,2,3,..$ is
   \bea
 \label{Sdalpha-answer-i}
S_{BG}&=&  \frac{L ^d}{2^d\pi ^{\frac{d}{2}}\beta^{\frac{d}{2 }}
   \lambda
^{\frac{d}{2}}}\left(\frac{d
   }{2}+1\right)
    \zeta \left(\frac{d
   }{2}+1\right),
   \eea
where $\lambda$ is a positive constant and $\zeta$ is the Riemann zeta function \cite{KV}, see Sect.\ref{sect:D4} for details.
The expression \eqref{Sdalpha-answer-i}
admits an analytical continuation to the complex $d$. We have the equality $S_{BH}=S_{BG}$ if $d=-\,4$ and $\lambda^2=3L^4/(64\,\pi ^3\,G)$. Therefore,  the entropy of the 4-dimensional Schwarzschild black hole coincides with the entropy of the Bose gas in $d=-\,4$ spacial dimensions. The extension of this result to the $D$-dimensional Schwarzschild black hole is considered in Sect.\ref{sect:D}. The Bose gas interpretation of the de Sitter entropy is discussed in Sect.\ref{sect:dS}.\\

The problem of the microscopic  origin of the Bekenstein-Hawking black hole entropy \cite{Bekenstein:1973,Hawking:1975vcx} 
has attracted enormous attention since the discovery of this entropy.
$D$-0 branes interpretation has been suggested  by Strominger and Vafa \cite{SV}. 't Hooft proposed to relate  black hole entropy with the entropy of thermally excited quantum fields in the vicinity of the horizon \cite{tHooft:1984kcu}. Note also  the recent searches (see  \cite{Chakravarty:2020wdm,Balasubramanian:2022gmo} and references therein)  for internal geometries that provide the entropy of a black hole. As was mentioned in \cite{Almheiri:2020cfm} this  resembles an old Wheeler's consideration of the black hole interior  as\ ''bag of gold''  \cite{Wheeler}. The  quantum model  corresponding to  black hole in  spacetime with topology $AdS_2\times S^8$ considered   recently by Maldacena \cite{Malda} satisfies the third law. 
\\

\section{Schwarzschild Black Hole in $D=4$ and Bose Gas in $d=-\,4$
}\label{sect:D4}
   
   The metric of the Schwarzschild black hole in $D=4$ spacetime is 
   \be
\label{metric4}
ds^2 = -\left( 1- \frac{r_h}{r} \right) dt^2
+
\cfrac{dr^2}{1- \frac{r_h}{r}}
+ r^2 d\omega_{2}^2,
\ee
where $d\omega_{2}^2$ is the metric on 2-dimensional unit sphere,  $r_h=2M$, $M$ is the mass of the black hole.
The Hawking temperature of Schwarzschild black hole in D=4 is
\be
T= \frac{1}{8 \pi G M},
\ee
where $M$ is the mass of the black hole. The Bekenstein-Hawking entropy and the free energy are
\bea
&S =  \cfrac{1}{16 \pi G T^2}, \\
&F_{BH} =  \cfrac{\beta}{16\pi G}.\label{FBH4}
\eea
The entropy  goes to infinity as $T\to 0$ and we see a violation of the third law in Planck's formulation.

  Now consider the non-relativistic Bose gas in d-spatial dimensions \cite{LLV,VZ} with the grand potential
  \bea
\label{2.2mm}
F_{BG}&=&\frac1\beta \sum _{\tiny{\begin{array}{cc}
    &k_i=  2\pi n_i/L \\
    &i=1,...d\\
    &n_i=1,2,...  
\end{array}}}  \ln\Big(1-
e^{\beta\left(\mu - \,\lambda\,\vec k ^2\right)}\Big), \quad \vec k ^2=\sum _{i=1,...d}k_i^2, \quad \lambda>0
\eea
in the box of the volume $L^d$ and the chemical potential $\mu$.

 The continuum version for large $L$ is
\bea 
\label{fs}F_{BG}(\beta)=\frac{L^d}{ (2\pi)^d\beta}\int\,\ln
\left(1-e^{\beta\left(\mu - \,\lambda\,
\vec k^2\right)}\right)\, d^d k.
\eea 
In what follows we take $\mu=0$. Using spherical coordinates and integrating over spherical angles in \eqref{fs}, we get
\bea
\label{fsdalpha-angle}F_{BG}&=&
\frac{\Omega_{d-1}}{ \beta}\left(\frac{L}{2\pi}\right)^d\int _{0}^{\infty}\ln
\left(1-e^{- \beta\,\lambda\,
 k^2}\right)\,k^{d-1} dk,
 \eea
 where $\Omega_{d-1}=\frac{2\pi ^{d/2}}{\Gamma(d/2)}$. 
Making the change of variables $\beta\,\lambda \, k^2=x$
and using the representation 
\bea 
\label{repr}
\int _{0}^{\infty}\ln
\Big(1-e^{- x}\Big)\,x^{s-1} dx=
 -\,\Gamma (s)\,\zeta(s+1),
\eea
valued for $\Re s>1,$
we get
\bea
 \label{fsdalpha-answer}
F_{BG}&=& 
    -\,\frac{L^d}{2^d\pi^{\frac{d}{2}}\beta^{\frac{d}{2 }+1}
   \lambda
^{\frac{d}{2}}}\,
    \zeta \left(\frac{d
   }{2}+1\right).
   \eea
   The expression \eqref{fsdalpha-answer} is obtained  for $\Re \,d>2,$ however $ \zeta (s)$ can be analytically extended to the hole complex plane as a meromorphic function with  the only simple pole at $s=1$.
   So we can use the formula \eqref{fsdalpha-answer} for complex $d$.
   The entropy is
   \bea
 \label{Sdalpha-answer}
S_{BG}&=& \frac{L ^d}{2^d\pi ^{\frac{d}{2}}\beta^{\frac{d}{2 }}
   \lambda
^{\frac{d}{2}}}\left(\frac{d
   }{2}+1\right)
    \zeta \left(\frac{d
   }{2}+1\right).
   \eea

 Equalizing \eqref{fsdalpha-answer} and the free energy of the 4-dimensional Schwarzschild black hole \eqref{FBH4} 
 \be
 \label{FF} 
 F_{BH}(\beta)=F_{BG}(\beta),
 \ee
 we get
 \bea
 \label{dD}
   -\,\frac{L ^d}{2^d\pi ^{\frac{d}{2}}\beta^{\frac{d}{2 }+1}
   \lambda
^{\frac{d}{2}}} \,
    \zeta \left(\frac{d
   }{2}+1\right)= \frac{\beta}{16\pi G}.\eea
  To fulfill this relation we have to  assume
   \bea
   d&=&-\,4\\\lambda^{2} &=&-\,\frac{L^4}{256\,\pi ^3 \zeta (-1) \,G}.\eea
   Taking into account that $\zeta (-1)=-1/12$,
   we get
   \bea\frac{\lambda^2}{L^4} &=&\frac{3}{64\,\pi ^3  \,G},\eea
therefore, we obtain that the thermodynamics of the 4-dimensional Schwarzschild black hole is equivalent to the 
thermodynamics of the Bose gas in $d=-\,4$ spacial dimensions. We understand
the 
thermodynamics of the Bose gas in {\it negative} spacial dimensions  in the sense of the analytical continuation of the right hand site of \eqref{fsdalpha-answer}.

   \section{Higher Dimensional  Black Holes and  Bose Gas}\label{sect:D}
   Let us recall that 
   for $D$-dimensional Schwarzschild black hole, $D\geq 4 $,
   \begin{equation}
\label{metric}
ds^2 = -\left( 1- \frac{{r_h}^{D-3}}{r^{D-3}} \right) dt^2
+
\cfrac{dr^2}{1- \frac{{r_h}^{D-3}}{r^{{D-3}}}}
+ r^2 d\omega_{D-2}^2,
\end{equation}
Hawking temperature $T=1/\beta$
is
\bea T&=&\frac{D-3}{4\pi r_h},
\eea
where $r_h$ is the  radius of the horizon. 
The entropy and the free energy are 
\bea
\label{SD}
S_{BH}&=&
\frac1{4 \,G_D}\left(\frac{D-3}{4\pi } \,\frac{1}{T}\right)^{D-2}\Omega_{D-2},\\
\label{FD}
F_{BH}&=&
  \frac{(D-3)^{D-3}}{4\,G_D (4\pi)^{D-2} } \, \beta^{D-3} \,\Omega_{D-2}
 \eea
where $G_D$ is the Newton constant in D-dimentional spacetime. Entropy $S$ goes to $\infty$, when $T\to 0$, 
and 
we obtain a violation of the third law in Planck's formulation again.

The free energy of the perfect gas of quasi-particles with energy $\varepsilon(\vec k)$  is given by \cite{VZ} 
\bea
\label{2.1}
F(\beta)=\frac1\beta \sum _{\vec {k}} \ln\Big(1-e^{- \beta\,\varepsilon(\vec k)}\Big),
\eea
where $\beta>0$ is the inverse temperature and 
$\vec k=\{k_1,...k_d\}$ sums over all one particle states of the system in d-spatial dimension.
Let us consider the free energy \eqref{2.1} with 
\bea
\label{epsilon-s}
\varepsilon_\alpha(\vec k)=
\lambda \left(\vec k^2\right)^{\alpha/2},
\eea
$\lambda$ is a positive
 dimensional constant.

In the continuum version we start from
\bea
\label{fsdalpha-d}F_{BG}&=&
\frac{\Omega_{d-1}}{ \beta}\left(\frac{L}{2\pi}\right)^d\int _{0}^{\infty}\ln
\left(1-e^{- \lambda\,\beta\,
 \,k^\alpha}\right)\,k^{d-1} dk.
 \eea
Making the change of variables $\lambda \beta k^\alpha=x$ and using representation \eqref{repr} we get
\bea
 \label{dDIV}
 F_{BG} = -\,\left(\frac{L}{2\pi}\right)^d\frac{2 \pi ^{d/2}}{d\Gamma(d/2)}\left(\frac{1}{\beta
   }\right)^{\frac{d}{\alpha }+1}\left(\frac{1}{\lambda
   }\right)^{\frac{d}{\alpha }}
   \Gamma \left(\frac{d}{\alpha
   }+1\right) \zeta \left(\frac{d
   }{\alpha }+1\right).
 \eea

Supposing that this free energy is equivalent to the D-dim  dimensional Schwarzschild free energy we get
 \bea
 -\,\frac{1}{d}\,\left(\frac{L}{2\pi}\right)^d\,\frac{2 \pi ^{d/2}}{ \Gamma(d/2)}\left(\frac{1}{\beta
   }\right)^{\frac{d}{\alpha }+1}\left(\frac{1}{\lambda
   }\right)^{\frac{d}{\alpha }}
   \Gamma \left(\frac{d}{\alpha
   }+1\right) \zeta \left(\frac{d
   }{\alpha }+1\right)=\frac{(D-3)^{D-3}}{4 G_D (4\pi)^{D-2} } \,\beta^{D-3}\,\frac{2 \pi ^{\frac{D-1}{2}}}{\Gamma(\frac{D-1}{2})}.\nn\\\label{d-D}\eea
    To equalize the powers of $\beta$ in both sizes of 
 \eqref{d-D} we take  
 \be\label{d-alpha}
 d=-\,(D-2)\,\alpha,  \ee
 and we get
\bea\label{dtoD}
 F_{BG} = -\,\left(\frac{2}{L}\right)^{(D-2)\,\alpha}\,\frac{ \pi ^{\frac{(D-2)\alpha}{2}}}{\Gamma(1-\frac{(D-2)\alpha}{2})}\,\beta
   ^{D-1}\,\lambda
   ^{D-2}
   \Gamma \left(3-D\right) \zeta \left(
   3-D\right).
 \eea
  Here and below we understand $D$ as $D+0$. 
Using the functional relation
\bea
\label{Gamma3D}
\Gamma(3-D)\,\zeta(3-D)=\frac{\zeta(D-2)}{2^{D-2} \pi^{D-3}\sin(\frac{\pi (D-2)}{2})},
\eea
we obtain necessary and sufficient conditions for the existence of $0<\lambda<\infty$ solving equation
 \eqref{d-D}
 \be
 \label{ineq-i}
 0<- \,\Gamma\left(1-\frac{(D-2)\alpha}{2}\right)\sin\left(\frac{\pi (D-2)}{2}\right)<\infty.\ee
\\
Since $D$ is a natural number, $\sin(\frac{\pi (D-2)}{2})$ 
takes three values:
\bea\label{3cases}
\sin(\frac{\pi (D-2)}{2})=\left\{\begin{array}{ccll}
    1 & \mbox{for} & D=4k+3,&\quad k=1,2,3,...\\
    0 & \mbox{for} & D=2k,&\quad k=2,3,4,...\\
    -1 &  \mbox{for} & D=4k+1,&\quad k=1,2,3,...
\end{array}\right.\,.\eea

Let us consider three cases \eqref{3cases} separately.
\begin{itemize}
 \item For $D=2k$ one has to assume, that the zeros of the sine must be compensated by the poles of the gamma function,
so we get
\be
1-\frac{(2k-2)\alpha}{2}=-\,p,\quad p=0,1,2,...,\ee
and 
\be
\label{alpha}
\alpha=\frac{(p+1)}{k-1}, \quad \mbox{where}\quad
p=0,1,2,..., \quad k=2,3,4,...\ee
Using  the functional relation 
\bea \label{GG}\Gamma\left(\frac{(D-2)\alpha}{2}\right)\,\Gamma\left(1-\frac{(D-2)\alpha}{2}\right)&=&\frac{\pi}{\sin (\pi \frac{(D-2)\alpha}{2})}\eea 
we rewrite conditions \eqref{ineq-i} as
\be
\label{ineq-m}
 0<- \,\frac{\pi \sin(\frac{\pi (D-2)}{2})}{\Gamma(\frac{(D-2)\alpha}{2})\sin (\pi \frac{(D-2)\alpha}{2})}<\infty.\ee
For positive $\alpha$  the 
 necessary and sufficient conditions 
\eqref{ineq-m} becomes
\be
\label{ineq-mm}
 0<- \,\frac{\sin(\frac{\pi (D-2)}{2})}{\sin (\pi \frac{(D-2)\alpha}{2})}<\infty.\ee
 For $D=2k$ and $\alpha$ as in \eqref{alpha}  the ratio of sines in \eqref{ineq-mm}
 becomes equal to $(-1)^{k-p-1}$,
and we get
\be
p=k+2l+1,\quad l=0,1,2.\ee
Therefore 
 our solution in this case is 
\be
D=2k,\quad \alpha=
\frac{(k+2l)}{k-1}.\ee
\item For $D=4k+1$, since  $\sin(\pi (4k-1)/2) =-1$, to have \eqref{ineq-i} we must require  
\be
 \label{ineq-k}
 0<\Gamma\left(1-\frac{(4k-1)\alpha}{2}\right)<\infty.\ee
\begin{itemize} 
\item For $\alpha<0$ the inequalities \eqref{ineq-k} are satisfied.
\item For $\alpha>0$ the inequalities \eqref{ineq-k} are 
satisfied only for 
\be 
\label{dom1}\frac{ 4 r}{4k-1}<\alpha<\frac{ 2(2r+1)}{4k-1}.
\ee

\end{itemize}

\item For $D=4k+3$, since  $\sin(\pi (4k-1)/2) =1$, to have \eqref{ineq-i} we must require  
\be
 \label{ineq-kk}
 0<-\,\Gamma\left(1-\frac{(4k-1)\alpha}{2}\right)<\infty.\ee
\begin{itemize} 
\item For $\alpha<0$ the inequalities \eqref{ineq-kk} are not satisfied.
\item For $\alpha>0$ the inequalities \eqref{ineq-kk}
only holds for 
\be 
\label{dom2}\frac{2(2 r+1)}{4k+1}<\alpha<\frac{4( r+1)}{4k+1}.
\ee

\end{itemize}
\end{itemize}

To summarize, we get the following 4 series of solutions

\begin{table}[h]
\begin{tabular}{|c|c|c|}
\hline
$D$                & $d$         & $\alpha$               \\ \hline
$D=4k+1,\quad k=1,2,3...$ & $d=(4k-1)|\alpha|$ & $\alpha=-1,-2,3,...$                                                                               \\ \hline
$D=4k+1,\quad k=1,2,3...$ & $d=-(4k-1)\alpha$   & $\,\,\,\,\,\frac{ 4 r}{4k-1}<\alpha<\frac{ 2(2r+1)}{4k-1},\quad r=0,1,2,...$                        \\ \hline
$D=4k+3,\quad k=1,2,3...$ & $d=-(4k+1)\alpha$   & \multicolumn{1}{l|}{$\frac{2(2 r+1)}{4k+1}<\alpha<\frac{4( r+1)}{4k+1},\quad r=0,1,2...$} \\ \hline
$D=2k,\qquad k=2,3,4...$  & $d=-2(k-1)\alpha$   & \multicolumn{1}{l|}{$\alpha=\frac{k+2n}{k-1},\quad  \,n=0,1,2,..$}                        \\ \hline
\end{tabular}
\caption{}\label{tab:equiv}
\end{table}

\section{De Sitter and Bose Gas}\label{sect:dS}
Let us remind that temperature,  entropy and the free energy (density ?) of the dS spacetime,
\be
 ds^2=-\left(1-\frac{r^2}{\ell^2}\right)dt^2+\left(1-\frac{r^2}{\ell^2}\right)^{-1}dr^2+r^2 d\omega_{D-2}^2,
\ee
are 
\bea
T&=&\frac{1}{2\pi\ell},
\\
S_{dS}&=& \frac{1}{4 G_N}\ell^{D-2}\Omega _{D-2}=\frac{\beta^{D-2}}{4G_D(2 \pi )^{D-2}}\Omega _{D-2},\\
F_{dS}&=&\frac{\beta ^{D-3}}{4G_D(2\pi)^{D-2}(D-3)}\Omega _{D-2},\eea
where $D\geq 3$. 

Note that \be
F_{dS}(\beta)=a(D)F_{BH}(\beta),\ee
where
\be a(D)=
\frac{2^{D-2}}{(D-3)^{D-2}}.\ee
 Therefore, the equation $F_{BG}=F_{dS}$ is reduced to equation $F_{BG}=a(D)F_{BH}$. Since $0<a(D)<\infty $
 we obtain the same solutions for $D,d$ and $\alpha$  as  presented in Table \ref{tab:equiv}.

\section{Discussion}
In this note we have shown that the thermodynamics of Schwarzschild black holes and de Sitter spacetime is equivalent to the thermodynamics of a Bose gas in spaces of negative dimensions.

\begin{itemize}
\item Equivalence is established using an analytic extension of the Riemann zeta function representation for the entropy of the Bose gas. We expect  a similar correspondence holding for other black hole solutions  and models of quantum statistics
\item 
Life in  negative dimensions seems to be more than just a fantasy. Some mathematically well-defined formulas allow for this exciting possibility. The Newton/Coulomb potential in d-dimensional space looks like $1/r^{d-2}$ and vanishes at large distances  for $d>2$. If $d$ is negative, the potential grows at infinity and  provides the confinement of matter in a trap, similarly to the de Sitter spacetime. 
\item Note that the functional relation for the Riemann zeta function
connects  negative and positive dimensions $d$.
Many properties of the Riemann zeta function, including the functional relations,  also hold for  Dirichlet  and more general $L$-functions. Partition functions of the bosonic string can be  expressed in terms of the $L$-functions
of a motive \cite{IVI,IVII}. Further generalizations are considered in  the Langlands program discussed in \cite{Witten,Arefeva:2007suy}. In \cite{Arefeva:2007suy} quantization of  the Riemann zeta function is considered. In \cite{Dragovich:2007wb} 
the zeta function string model has been proposed. Negative dimensions in a different context are discussed in 
\cite{Manin,Maslov}. 
Negative dimensions \cite{Anastasiou:1999ui}  in the spirit of non-integer dimensions \cite{tHooft:1972tcz}
are used for loop integrals.
\item We have discussed the correspondence between the black holes entropy in positive dimensions and Bose gas in negative dimensions. However, one can also consider a correspondence between the spacetime of negative dimensions and Bose gas in positive dimensions. Writing the relation \eqref{d-alpha} in the form
$D=2- d/\alpha$ and taking $d$ and $\alpha$  positive, we get 
a negative dimensional spacetime $D$ consistent with \eqref{d-D}.
\end{itemize}

In conclusion,  there is an interesting equivalence between the entropy of Schwarzschild black holes and the entropy of a Bose gas, given in the Table \ref{tab:equiv}. We see that either the dimension of the nonrelativistic Bose gas is negative, or we are dealing with a Bose gas with a nonstandard kinetic term. It would be interesting to find a similar correspondence for other black holes or branes.

\section*{Acknowlegments}

We would like to thank D. Ageev, V. Berezin, V. Frolov, M. Khramtsov, K. Rannu, P. Slepov, A. Teretenkov, A. Trushechkin and V. Zagrebnov for fruitful discussions. This work is supported by  the Russian Science Foundation (project
19-11-00320, V.A. Steklov Mathematical Institute).


\begin{thebibliography}{100}
 \bibitem{BCH}
B.Bardeen, B.Carter and S.W.Hawking, The four laws of black hole mechanics, {\sl Commun. Math. 
Phys.} {\bf 31}, 161 (1973).

 \bibitem{Bekenstein:1973}
J.~D. Bekenstein, ``Black holes and entropy,''
 Phys. Rev. D
  {\bf 7} (Apr, 1973) 2333--2346.
  
\bibitem{Hawking:1975vcx}
S.~W. Hawking, ``{Particle Creation by Black Holes},''
{Commun. Math. Phys.
  {\bf43} (1975) 199--220}. [Erratum: Commun.Math.Phys. 46, 206 (1976)] %
  
  \bibitem{FN} V. Frolov and I. Novikov, Black hole physics: basic concepts and new developments
(Vol. 96) (2012), Springer Science, Business Media.

   \bibitem{Wald-book} R.M. Wald, General Relativity, University of Chicago Press (1984).
 



\bibitem{LLV} L. D. Landau, E. M. Lifshitz  Statistical Physics: Volume 5. – Elsevier, 2013. – v. 5.

\bibitem{Israel}
W.Israel, Third Law of Black-Hole Dynamics. A Formulation and Proof,  Phys. Rev. Lett. {\bf 57}, 397 (1986).


\bibitem{Wald}
R.Wald,  "Nernst theorem" and black hole thermodynamics,  Phys. Rev. {\bf D56}, 6467 (1997); 
\bibitem{Racz} I. Rácz,  "Does the third law of black hole thermodynamics really have a serious failure?." Classical and Quantum Gravity 17.20 (2000): 4353.
 \bibitem{Belgiorno:2002pm}
F.~Belgiorno and M.~Martellini,
``Black holes and the third law of thermodynamics,''
Int. J. Mod. Phys. D \textbf{13}, 739-770 (2004)
[arXiv:gr-qc/0210026 [gr-qc]].
\bibitem{Wreszinski:2007gu}
W.~F.~Wreszinski and E.~Abdalla,
``A Precise formulation of the third law of thermodynamics with applications to statistical physics and black holes,''
[arXiv:0710.4918 [math-ph]].



 \bibitem{Kehle:2022uvc}
C.~Kehle and R.~Unger,
``Gravitational collapse to extremal black holes and the third law of black hole thermodynamics,''
[arXiv:2211.15742 [gr-qc]].





\bibitem{KV} A. A. Karatsuba and S. M. Voronin. "The Riemann Zeta-Function "(de Gruyter, Berlin (1992).
\bibitem{SV} A. Strominger and C. Vafa, “Microscopic origin of the Bekenstein-Hawking entropy,” Phys. Lett. B 379 (1996) 99–104, arXiv:hep-th/9601029.




\bibitem{tHooft:1984kcu}
G.~'t Hooft,
``On the Quantum Structure of a Black Hole,''
Nucl. Phys. B \textbf{256}, 727-745 (1985)
\bibitem{Chakravarty:2020wdm}
J.~Chakravarty,
``Overcounting of interior excitations: A resolution to the bags of gold paradox in AdS,''
JHEP \textbf{02}, 027 (2021)
[arXiv:2010.03575 [hep-th]].
\bibitem{Balasubramanian:2022gmo}
V.~Balasubramanian, A.~Lawrence, J.~M.~Magan and M.~Sasieta,
``Microscopic origin of the entropy of black holes in general relativity,''
[arXiv:2212.02447 [hep-th]].


\bibitem{Almheiri:2020cfm}
A.~Almheiri, T.~Hartman, J.~Maldacena, E.~Shaghoulian, and A.~Tajdini, ``{The
  entropy of Hawking radiation},''
Rev. Mod. Phys.
  {\bf 93} no.~3, (2021) 035002, arXiv:2006.06872
  [hep-th].
  
\bibitem{Wheeler}
J.~Wheeler, ``{\it Relativity, groups and topology},'' edited by B. S. DeWitt
  and C. M. DeWitt. Gordon and Breach, New York, 1964 .

\bibitem{Malda}
J.~Maldacena,
``A simple quantum system that describes a black hole,''
[arXiv:2303.11534 [hep-th]].




\bibitem{VZ} V.A. Zagrebnov and J.-B. Bru. "The Bogoliubov model of weakly imperfect Bose gas." Physics Reports 350.5-6 (2001) 291-434.


\bibitem{IVI} I. V. Volovich,  "D-branes, black holes and SU ($\infty$) gauge theory." arXiv preprint hep-th/9608137 (1996): 41-48.
\bibitem{IVII} I. V. Volovich, “From $p$-adic strings to \'{e}tale strings”, Proc. Steklov Inst. Math., 203 (1995), 37–42 

\bibitem{Witten} S. Gukov and E. Witten, Gauge Theory, Ramification, And The Geometric Langlands
Program, Communications in Number Theory and Physics,  Vol. 1, p.1-236 (2007),  hep-th/0612073;
\bibitem{Arefeva:2007suy}
I.~Y.~Aref'eva and I.~V.~Volovich,
``Quantization of the Riemann Zeta Function and Cosmology,''
Int. J. Geom. Meth. Mod. Phys. \textbf{4}, 881-895 (2007)
[arXiv:hep-th/0701284 [hep-th]].
\bibitem{Dragovich:2007wb}
B.~Dragovich,
``Zeta strings,''
[arXiv:hep-th/0703008 [hep-th]].
\bibitem{Manin} Yuri Manin,  "The notion of dimension in geometry and algebra." Bulletin of the American Mathematical Society 43.2 (2006), 139-161.

\bibitem{Maslov}  V. P. Maslov, "Negative dimension in general and asymptotic topology." arXiv math/0612543 (2006).
\bibitem{Anastasiou:1999ui}
C.~Anastasiou, E.~W.~N.~Glover and C.~Oleari,
``Scalar one loop integrals using the negative dimension approach,''
Nucl. Phys. B \textbf{572}, 307-360 (2000)
[arXiv:hep-ph/9907494 [hep-ph]].

\bibitem{tHooft:1972tcz}
G.~'t Hooft and M.~J.~G.~Veltman,
``Regularization and Renormalization of Gauge Fields,''
Nucl. Phys. B \textbf{44}, 189-213 (1972)
doi:10.1016/0550-3213(72)90279-9
\end{thebibliography}
\end{document}